%% file: ms.tex
\pdfoutput=1
\documentclass[letterpaper, 10 pt, conference]{ieeeconf}  

\IEEEoverridecommandlockouts                              
\usepackage{graphicx}
\usepackage{subfiles}
\usepackage{epsfig} 
\usepackage{mathptmx} 
\usepackage{times} 
\usepackage{amsmath} 
\usepackage{amssymb}  
\usepackage{subfig}
\usepackage{hyperref}
\usepackage{float}
\usepackage[noadjust]{cite}

\title{\LARGE \bf
Design and Implementation of a Three-Link Brachiation Robot with Optimal Control Based Trajectory Tracking Controller
}

\author{
Shuo Yang$^{1,2}$, 
Zhaoyuan Gu$^{1,2}$, Ruohai Ge$^{1}$,
Aaron M. Johnson$^{2}$, Matthew Travers$^{1}$ and Howie Choset$^{1}$%
\thanks{$^{1}$Biorobotics Laboratory, The Robotics Institute, Carnegie Mellon University, Pittsburgh, PA 15213, USA
{\tt\small \{shuoyang, zhaoyuan, ruohaig, mtravers, choset\}@andrew.cmu.edu}}
\thanks{$^{2}$Department of Mechanical Engineering, Carnegie Mellon University, Pittsburgh, PA 15213, USA {\tt\small amj1@andrew.cmu.edu}}%
}

\begin{document}

\maketitle
\thispagestyle{empty}
\pagestyle{empty}

\begin{abstract}

This paper reports the design and implementation of a three-link brachiation robot. The robot is able to travel along horizontal monkey bars using continuous arm swings. We build a full order dynamics model for the robot and formulate each cycle of robot swing motion as an optimal control problem. The iterative Linear Quadratic Regulator (iLQR) algorithm is used to find the optimal control strategy during one swing. We select suitable robot design parameters by comparing the cost of robot motion generated by the iLQR algorithm for different robot designs. In particular, using this approach we show the importance of having a body link and low inertia arms for efficient brachiation.
Further, we propose a trajectory tracking controller that combines a cascaded PID controller and an input-output linearization controller to enable the robot to track desired trajectory precisely and reject external disturbance during brachiation. Experiments on the simulated robot and the real robot demonstrate that the robot can robustly swing between monkey bars with same or different spacing of handholds. 

\end{abstract}

\section{INTRODUCTION}\label{sec_intro}
\subfile{sections/sec_intro}

\section{RELATED WORK}\label{sec_related}
\subfile{sections/sec_related}

\section{ROBOT MODELING}\label{sec_model}
\subfile{sections/sec_model}

\section{TRAJECTORY GENERATION}\label{sec_gen}
\subfile{sections/sec_traj_gen}

\section{TRAJECTORY TRACKING CONTROLLER}\label{sec_ctrl}

\subfile{sections/sec_traj_ctrl}

\section{ROBOR DESIGN AND IMPLEMENTATION}\label{sec_exp}
\subfile{sections/sec_experiment}
\section{CONCLUSION}\label{sec_conclude}
\subfile{sections/sec_conclude}

\section*{ACKNOWLEDGMENT}
We would like to thank Alex Sun, Bo Tian and Weijia Gao for their help on building the robot hardware.

\bibliographystyle{IEEEtran}
\bibliography{sections/references}

\end{document}

%% file: sections/sec_intro.tex

Brachiation is a locomotion strategy employed by monkeys and many other primates. For certain species of gibbon, brachiation occupies 80\% of their movements \cite{birx2006encyclopedia}. Many biologists have studied this type of motion for decades \cite{fleagle1974dynamics,jungers1981preliminary,chang1997dynamic,michilsens2009functional}. To build a robot that can brachiate is challenging because the motion is highly dynamic, nonlinear, and underactuated. Therefore, a brachiation robot is a good platform to study and research new control strategies. 
Moreover, by comparing the motion of the robot to that of real animals, the knowledge we learn during the design of a brachiation robot can provide insights into how animals generate their behaviours and how to reproduce these behaviours on robotic systems. 

\begin{figure} 
    \centering
  \subfloat{%
       \includegraphics[width=0.51\linewidth]{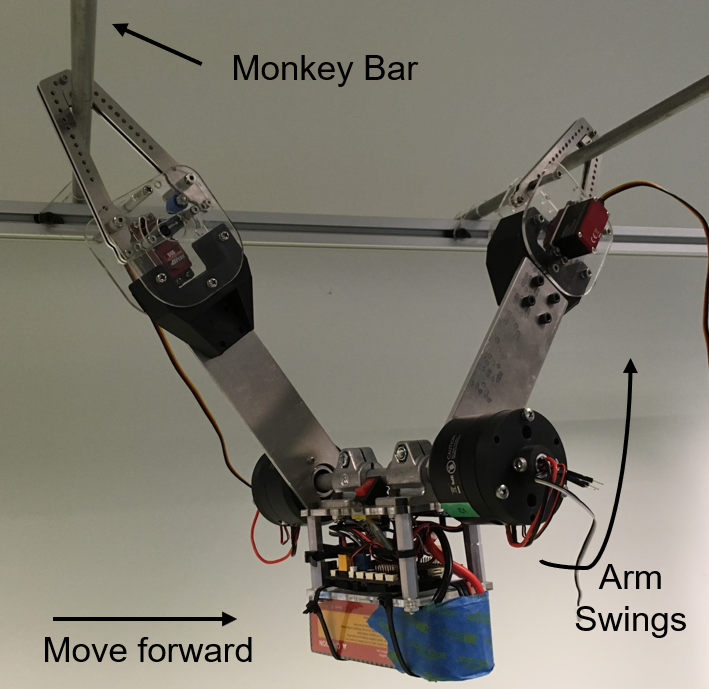}}
  \subfloat{%
        \includegraphics[width=0.517\linewidth]{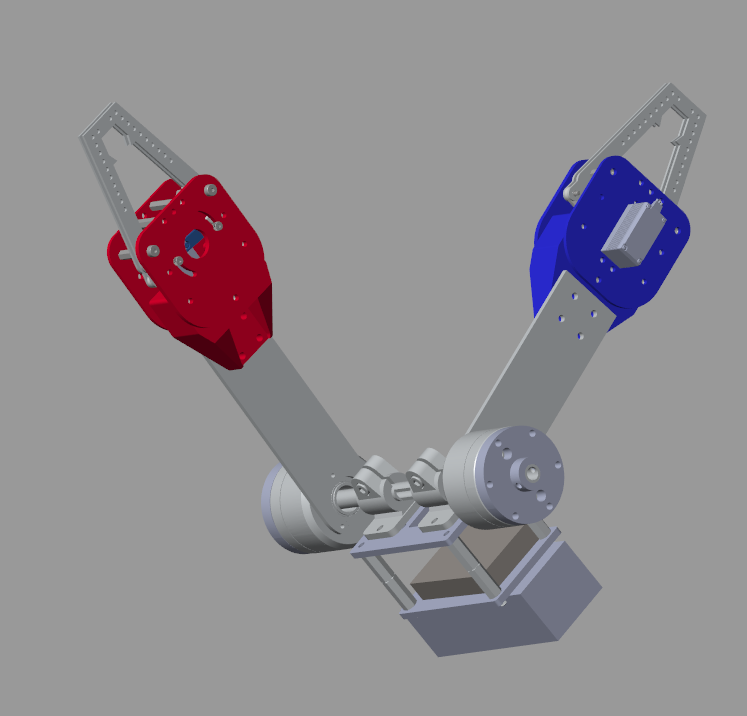}}
  \caption{(a) The brachiation robot hardware. (b) Simulink robot model}
  \label{fig:sec1_monkey}
\end{figure}

In this work we build a three-link brachiation robot, shown in Fig.~\ref{fig:sec1_monkey}, consisting of two arms and one body. During brachiation, the robot has three joints but only two of them are actuated. We formulate each brachiation swing motion cycle as an optimal control problem and solve the problem using the iterative Linear Quadratic Regulator (iLQR) algorithm \cite{li2004iterative}. The iLQR algorithm and a robot simulation form a testbed to consider different robot designs. Using the testbed, we can efficiently change robot physical parameters according to animal data presented in previous studies \cite{chang1997dynamic,michilsens2011pendulum} and use the final value of the objective function of the optimal control problem to evaluate design decisions. 

To robustly control the real robot hardware, we propose a trajectory tracking controller that tracks both the configuration space task and the end effector space task. The end effector space task controller is derived using input-output linearization \cite{khalil2002nonlinear}. The controller is validated in simulation and on real robot hardware. With the proposed controller, the robot can swing between monkey bars with different spacing of handholds and execute multiple swings continuously.

This paper has three contributions:
\begin{enumerate}
    \item The modeling, design, and implementation of a three-link robot that achieves robust brachiation motion.
    \item A trajectory tracking controller for the brachiation robot that enables robust tracking of control tasks despite the unactuated swinging joint.
    \item A design methodology to select robot parameters to achieve energy efficient brachiation, using the optimal control objective function as an evaluation metric for robot design, that shows the importance of the body link and low inertia arms for brachiation.
\end{enumerate}

The paper is organized as follows. Section \ref{sec_related} provides a literature review. Section \ref{sec_model} presents the modeling of the system dynamics. Section \ref{sec_gen} describes the iLQR algorithm for trajectory generation. Section \ref{sec_ctrl} presents the trajectory tracking controller. Section \ref{sec_exp} discusses robot design, implementation and experiments. Section \ref{sec_conclude} concludes the paper and discusses future work.

%% file: sections/sec_related.tex
\subsection{Brachiation}
During one continuous-contact brachiation cycle, the animal or robot can be simply viewed as a suspended pendulum \cite{parsons1977energetics}, but more complex behaviours are also observed. One of the earliest studies of animal brachiation is presented in \cite{fleagle1974dynamics}. It concludes that complicated energy exchanges among trunk, legs, and arms are involved during brachiation. For example, a siamang moves its center-of-mass (COM) near the center of rotation by stretching its legs and swinging its arms. Following works suggest that fast brachiation and ricochetal brachiation rely on the rotation of the body to gain additional momentum \cite{bertram2001mechanical} and that the body can also help reduce collision energy loss \cite{usherwood2003understanding}. Moreover, a recent anatomy study of the gibbon forelimb indicated that muscles in the forelimb and shoulder regulate the body COM movement to achieve energy efficient brachiation \cite{michilsens2009functional}. 

\subsection{Brachiation Robots}
Many brachiation robots have been built in the past.
The pioneering work on brachiation robots was conducted by researchers in Fukuda's group \cite{fukuda1991brachiation}. Their brachiator II \cite{fukuda1996motion} was a two-link (one for each arm), single-actuator brachiation robot, which did not have a body. The control strategy of this robot was based on a heuristic model. 
The Brachiator III \cite{nakanishi1998experimental} looks a lot like a real monkey with two arms, a body and a pair of legs. But the robot moved very slow so it just used its legs to gain initial velocity rather than explicitly consider the body and leg swings in the controller design. 
The work presented in \cite{de2007control} discussed the modeling of a three-link brachiation robot and presented simulation results. 
\cite{mazumdar2009mag} designed a two-link brachiation robot for bridge inspection. \cite{lo2017model} built three-link robot but the arms of their robot were too short to generate dynamical effects, thus in their work arms were neglected in the robot model. The recent ``Tarzan'' robot \cite{farzan2018modeling} was mainly focused on how to traverse along a flexible cable, while the robot was still designed to have two-link structure. 

\subsection{Trajectory Optimization \& Control}
For a given system dynamics model with a control input, trajectory optimization aims to find a control and state sequence that optimizes a cost function for this system during a certain period of time. The problem can be solved by differential dynamic programming (DDP) \cite{mayne1966second}. A variation of DDP is iterative Linear Quadratic Regulator (iLQR) \cite{li2004iterative} which ignores second order terms in some DDP steps to speed up the algorithm. iLQR has been used to optimize trajectories for high dimensional dynamic systems such as humanoid robot \cite{tassa2014control}. 

Although a trajectory optimization result contains a control sequence, the sequence is rarely used directly to control the real robot because the model used in the optimization algorithm inevitably differs from the true dynamic system in the physical world. Feedback control must be used to compensate model mismatch. A common control scheme for a nonlinear system is dynamic feedback linearization \cite{charlet1989dynamic}, which is widely used in fixed-base manipulator control \cite{lynch2017modern}. For a underactuated system, only partial feedback linearization can be achieved \cite{spong1995swing}. Partial feedback linearization is also called input-output linearization \cite{khalil2002nonlinear} or zero dynamics \cite{isidori2013nonlinear}. An extension of zero dynamics, with the name hybrid zero dynamics, is popular in the bipedal walking robot control community \cite{westervelt2003hybrid}.

%% file: sections/sec_model.tex
\begin{figure} 
    \centering
  \subfloat{%
       \includegraphics[width=0.6\linewidth]{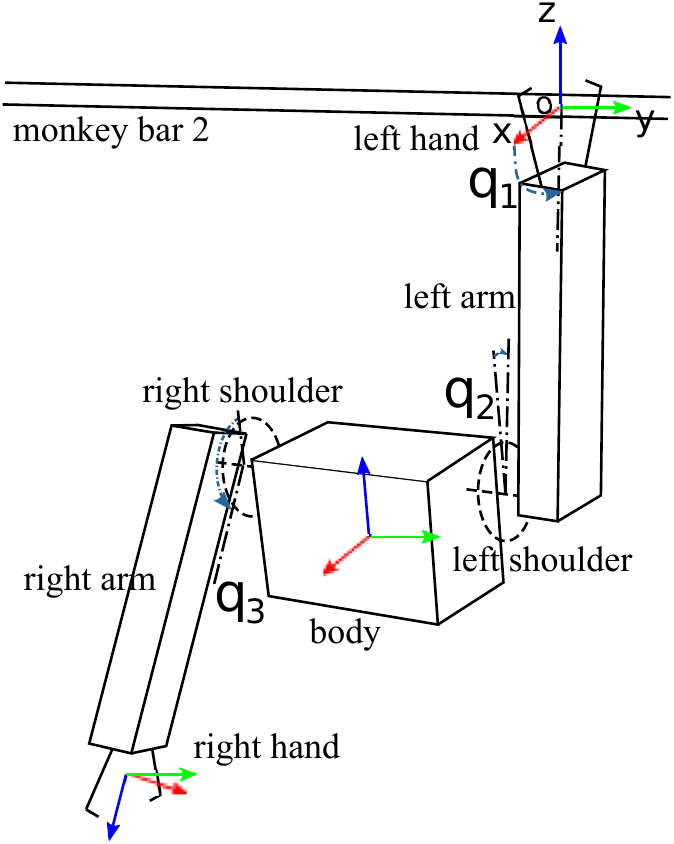}}\label{fig:sec_model_robot_o}
    \hfill
  \subfloat{%
        \includegraphics[width=0.8\linewidth]{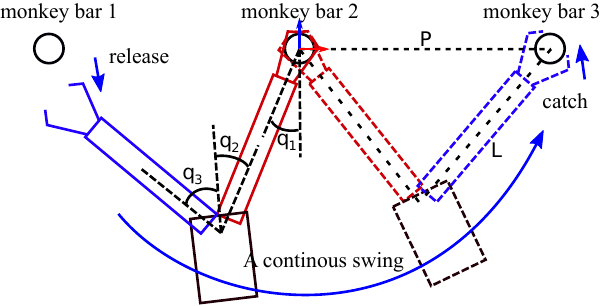}}\label{fig:sec_model_robot_i}
  \caption{(a) The model of the robot. (b) Illustration of one continuous swing cycle.}
  \label{fig:sec_model_robot}
\end{figure}

Based on the discussion of the importance of the body presented in biological brachiation studies, we built a three-link brachiation robot with one body and two arms linked by shoulder joints. Each arm has a gripper hand. The robot is shown in Fig. \ref{fig:sec1_monkey}. We will discuss its design parameters selection in Section \ref{sec_exp}. In this section, to simplify the robot modeling, we treat the robot as a collection of three connected rigid cuboids, Fig.~\ref{fig:sec_model_robot}(a).

During one continuous-contact brachiation cycle, without loss of generality, we assume the left hand is the holding hand and the right hand is the swing hand. At the start of the swing cycle, the left hand of the robot hangs on bar 2, and the robot releases the right hand gripper from bar 1. The right hand then swings forward to reach and catch bar 3, as illustrated in Figure \ref{fig:sec_model_robot}(b). 
After releasing the right hand, control torques need to be applied to the shoulder motors so that the right hand reaches the next bar by the end of the swing cycle. Then the right hand gripper must close at the correct time and location to catch the bar to complete one swing cycle. After finishing the current cycle, the robot hangs on bars 2 and 3, then the roles of two hands switch. The next cycle goes in the same way. During one cycle, the robot can be modeled as a three-link manipulator whose base link is at the center of bar 2 and the end effector frame locates at the center of right hand gripper. Figure \ref{fig:sec_model_robot}(a) shows the frame locations and joints of the manipulator. The rotation between the bar and the left arm forms joint $q_1$ and the two shoulder motors give joint $q_2$ and $q_3$. These joints are modeled as 1-DOF revolute joints without rotation limitations. It should be emphasised that although the robot has three joints, the joint $q_1$ is underactuated. This part of the model aligns with animal observations because, during brachiation, the wrist cannot generate enough torque to affect the motion.

\subsection{Dynamic Modeling}
Following this setup, the dynamic model of the robot is defined as

\begin{equation}\label{eqn:robot_dyn}
   M(q)\ddot{q} + C(q,\dot{q}) + G(q) = 
Bu 
\end{equation}
where 
$$
q = \begin{bmatrix}
q_1 \\ q_2 \\ q_3
\end{bmatrix} \ \ \ \
\dot{q} = \begin{bmatrix}
\dot{q}_1\\ \dot{q}_2\\ \dot{q}_3
\end{bmatrix} \ \ \ \
\ddot{q} = \begin{bmatrix}
\ddot{q}_1\\\ddot{q}_2\\\ddot{q}_3
\end{bmatrix}
\ \ \ \ 
B = \begin{bmatrix} 
0 & 0 \\ 
1 & 0 \\
0 & 1
\end{bmatrix}
\ \ \ \ 
u = \begin{bmatrix}
\tau_2 \\
\tau_3
\end{bmatrix}
$$
in which the state of the system $q$  consists of joint angles, $u$ is the control torque generated from shoulder motors, and $M(q)$, $C(q,\dot{q})$ and $G(q)$ are terms related to dynamic properties of the robot, the derivation of which follows standard rigid body dynamics, e.g.~\cite{lynch2017modern}. Even though motion happens only in the X-Z plane, we derive the model in 3D space for completeness. 

We define several important terms. The home configuration or the configuration of the robot when $q = 0$ is defined when the robot hangs on one bar with two arms together. It is easy to see when the two hands of the robot hang on different bars, angles in $q$ can be determined by the distance between bars and the length of arm. In Figure \ref{fig:sec_model_robot}(b), if the position of bar 3 relative to bar 2 is $p=(p_x, p_z)$, and the robot arm length is L, then $q_1 = \sin^{-1}(\frac{|P|}{2L})+\tan^{-1}(p_z/p_x)$, where $|P|^2 = q_x^2 + q_y^2$. $q_2$ can be freely chosen because body can rotate freely even if hand positions are fixed on bars. And $q_3 = \sin^{-1}(\frac{|P|}{2L}) - 2\pi + q_2$. The angles form ``final configuration'' $q_T$. Similarly we can define ``initial configuration'' $q_0$.  Since angle $q_2$ represents the offset between the central axes of the left arm and the body, we call it ``offset angle''. We follow the biological observation discussed in \cite{bertram2001mechanical} that during brachiation the body tries to extend as much as possible to move overall COM downward to set initial and final offset angles both to 0.

%% file: sections/sec_traj_gen.tex
To execute one brachiation motion cycle, the robot first hangs on the bars with the initial configuration $q_0$. Then it starts a timer from $0$ to $T$. For any time instance $t \in [0, T]$, the robot uses a controller $u_t \in \mathbb{R}^2$ to apply control torques to joint 2 and joint 3. At time instance $T$, the robot reaches a desired configuration $q^*_T$. In order for the robot to minimize motion errors as well as energy consumption, we can formulate the process as an optimal control problem.


\subsection{Problem Formulation}
We define system state $x = [q \ \ \  \dot{q}]^T$, so 
$$
\dot{x} = 
\begin{bmatrix}
\dot{q} \\
\ddot{q}
\end{bmatrix}
= 
\begin{bmatrix}
\dot{q} \\
M(q)^{-1}(Bu - C(q,\dot{q}) - G(q)) 
\end{bmatrix}
=
f(x,u)
$$
is the differential equation derived from the system dynamics. We can discretize it to get 
$$x_{t+1} = x_t + f(x_t, u_t)\Delta t$$
where the time difference between $t+1$ and $t$ is a predetermined small number $\Delta t$. Then a corresponding discrete control sequence can be defined as $\textbf{u} = \{u_0, u_1, ..., u_i,..., u_{Tn}\}$ where $i = 0...T_n$ and $T_n = T/\Delta T$. Thus $x_0  =  [q_0 \ \ 0 \ \ 0 \ \ 0]^T$ and $x^*_T  =  [q^*_T \ \ 0 \ \ 0 \ \ 0]^T$.

We define the following optimization problem
\begin{align}
    \mathop{\min}_{\textbf{u}}\ \  & J = (x_T-x_T^*)^TQ_f(x_T-x_T^*) + \sum_{i=0}^{T} (x_i^TQx_i + u_i^TRu_i) \label{eqn:problem}\\
    \text{subject to}\ \  & x_{i+1} - x_i= f(x_i, u_i)\Delta t \ \ \forall i=0...T_n \\
                              & x_0 = x(0)
\end{align}
in which objective function J contains two parts. The first part is the distance between the final configuration $x_T$ and the desired final configuration $x_T^*$. The second part is designed to minimize joint velocities and energy consumption of the control sequence. Notice that to achieve this control goal, we only assign nonzero values to the last three diagonal entries of matrix $Q$. This objective function represents a minimum collision energy loss strategy \cite{usherwood2003understanding} because ideally the robot will reach $q^*_T$ with zero velocity so that right hand gripper can gently catch the next bar.  

\subsection{iterative  Linear  Quadratic  Regulator}
iLQR is chosen to solve Problem \ref{eqn:problem} because it is a locally iterative method and very efficient. Although iLQR is a local method, in our case we only need the local minimum around an initial zero control sequence. The algorithm initially starts with zero control sequence $\textbf{u}^0$ (where superscript $j=0$ indicates iteration number) and iteratively refine this control sequence by
$$
\textbf{u}^{j+1} = \textbf{u}^{j} + \delta \textbf{u} = \textbf{u}^{j} + \textbf{k}^{j}  + \textbf{K}^{j}\delta x^{j}
$$
where $\delta x^{j}$ is the difference of running states before and after applying control updates in the previous iteration. $\textbf{k}^{j}$ and $\textbf{K}^{j}$ are two sequences of gains that each has $T_n+1$ elements. The derivation of expressions of $\textbf{k}^{j}$ and $\textbf{K}^{j}$ can be found in~\cite{tassa2014control}.

The algorithm goes as follows. In iteration 0, starting from $\textbf{u}^0$ and initial state $x_0$, the algorithm first runs a forward pass using system dynamics to get $\textbf{x}^0$, a sequence of states that are the result of control sequence $\textbf{u}^0$. Then a backward pass gives $\textbf{k}^{0}$ and $\textbf{K}^{0}$. Since there is no previous iteration, $\delta x^{0}$ is a all-zero sequence, and $\textbf{u}^{1} = \textbf{u}^{0} + \textbf{k}^{0}$. Now we move on to iteration 1, $\textbf{x}^1$ is generated by a forward passing using $\textbf{u}^1$ and initial state $x_0$ and $\delta x^{1} = \textbf{x}^1 - \textbf{x}^0$. Then we do backward pass again to get $\textbf{k}^{1}$ and $\textbf{K}^{1}$, and update $\textbf{u}^{2}$ to be $\textbf{u}^{2} = \textbf{u}^{1} + \textbf{k}^{1} + \textbf{K}^{1}\delta x^{1}$. Iterations continue, and during each forward pass the new value of the objective of the optimization $J$ is recorded. The algorithm stops when the value of $J$ at iteration $j+1$ does not change significantly compared to that in iteration $j$ (up to a desired tolerance). The $\textbf{u}^{j+1}$ in the last iteration is considered as the optimal control sequence $\textbf{u}^*$.

%% file: sections/sec_traj_ctrl.tex
The controller implemented on the robot hardware does not directly use the solved optimal control sequence $\textbf{u}^*$ to fit $u(t)$, but instead uses a trajectory tracking feedback controller.
\subsection{Controller Design}
We propose the following control strategy during trajectory tracking control:
$$
u(t) = u_{config}(t) + \alpha u_{task}(t) 
$$

The first part $u_{config}(t)$ tries to achieve configuration space task, namely letting the actuated joints ($q_2$ and $q_3$) of the robot track the desired joint position and velocity using cascaded PID controllers. For each individual dimension ($i=2,3$):
$$
u_{i,config}(t) = \text{PID}( \text{PID}(q_i^d(t) - q_i(t), \theta_{pos})+\dot{q}_i^d(t) - \dot{q}_i(t), \theta_{vel})
$$

where $o = \text{PID}(e, \theta)$ denotes a PID controller with parameter $\theta$ that takes in an error signal $e$ and outputs a control signal $o$. At each time instance, $q_i^d(t)$ and $\dot{q}_i^d(t)$ are interpolated from reference trajectories. The reference trajectories are pre-computed on the high-level computer by forward passing the optimal control sequence $\textbf{u}^*$ through the robot dynamics model and record all joint positions and velocities. $q(t)$ and $\dot{q}(t)$ are feedback signals generated by motor encoders.

The second part of the controller $u_{task}(t)$ provides additional control inputs to achieve end effector space task. If joint $q_1$ accumulates tracking errors due to disturbances, even though joint angles of $q_2$ and $q_3$ track their references perfectly, the end effector position can still differ from the desired value. Since the system is underactuated, the tracking error of joint $q_1$ can only be eliminated by the body and arm swings which are regulated by $u_{task}(t)$.  From $q^d_i(t)$, we can calculate the position of right hand $p^d(t) = FK(q^d(t))$ using the forward kinematics of the manipulator.
At each time instance $t$ during swinging, we can define the end effector tracking error $y(t)$ as
$$
y(t) = h(t,q) = p^d(t) - FK(q(t))
$$

So the end effector space task is keeping $y(t)$ asymptotically stable at the origin. To achieve this control task we use input-output linearization \cite{khalil2002nonlinear}. For system 
\begin{align*}
    \dot{q}  &= \dot{q} \\
    \ddot{q} &= M(q)^{-1}\left(-C(q,\dot{q}) - G(q) + Bu\right)   
\end{align*}
with output $y = p - FK(q)$. It can be shown that the system has relative degree 2, because the state feedback control 
\begin{align*}
 u_{task} = &\left(-\frac{\partial FK}{\partial q}M^{-1}B\right)^{-1} \\
 & \left(v -\ddot{p}-\frac{\partial FK}{\partial q}M^{-1}(C(q,\dot{q})+G(q))+
\frac{\partial}{\partial q}(\frac{\partial FK}{\partial q}\dot{q})
\dot{q}\right)
\end{align*}
reduces the input-output map to
$
\ddot{y} = v
$. If we further choose $v = -K_p y - K_d \dot{y}$, we get $\ddot{y} = -K_p y - K_d \dot{y}$. Here $K_p$ and $K_d$ are positive definite matrices with proper dimensions. This output dynamics can be theoretically asymptotically stabilized \cite{khalil2002nonlinear}. However, since $B$ is not a square matrix, we can only take pseudo-inverse of $\frac{\partial FK}{\partial q}M^{-1}B$.

\begin{figure} 
    \centering
  \subfloat{%
       \includegraphics[width=\linewidth]{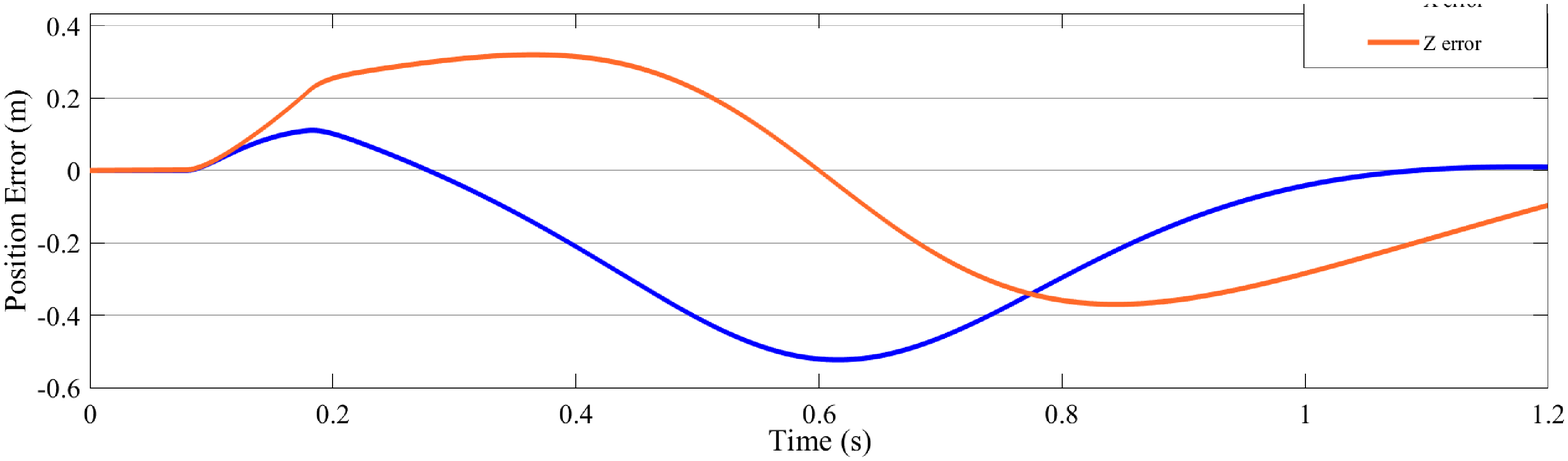}}
    \hfill
  \subfloat{%
        \includegraphics[width=\linewidth]{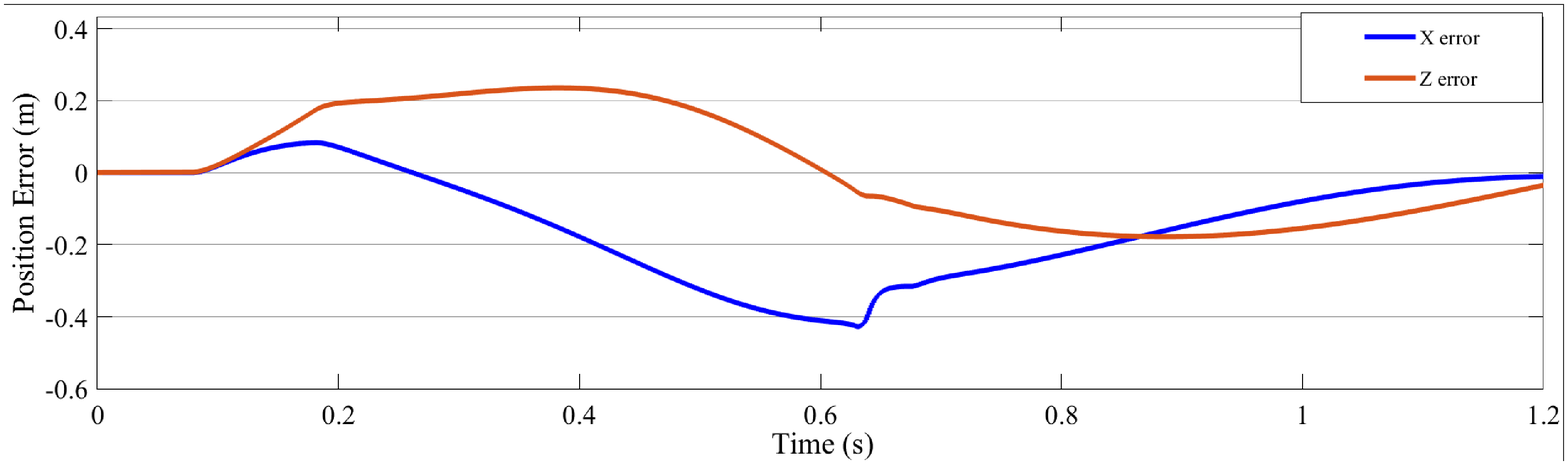}}
  \caption{(Top) Controller performance under disturbance with only $u_{config}$.  (Bottom) Controller performance with both $u_{config} + \alpha u_{task}$. The additional control term drives the error to close to zero in both $x$ and $z$.}
  \label{fig:control_compare}
\end{figure}

\subsection{Simulation Result}
We evaluate the proposed controller performance using the simulated robot presented in \ref{sec_exp_sim}. Our focus is to use $u_{task}$ to further reduce end effector position errors when the robot experiences disturbances. The simulation experiment uses a 9kg robot with 2kg arm weights executes brachiation using this controller. During the 1.2s motion cycle, the right hand experience a 20Nm drag force along positive z direction from 0.078s to 0.178s after swing starts (to simulate potential collisions between the hand and a bar). We first control the robot with only $u_{config}$, and then run the experiment with the proposed controller.

Figure \ref{fig:control_compare} shows the comparison of end effector position errors in the two cases. Plot (a) shows that the disturbance makes the end effector position quickly accumulate a large displacement and the final position of the end effector has a $0.1m$ error in z direction. With the additional control term $u_{task}$, shown in plot (b), the maximum position error changes from $0.52m$ to $0.42m$, and the final end effector position error is reduced to $0.02m$.




%% file: sections/sec_experiment.tex
This section discusses the design and implementation of the robot. We first report the simulated robot setting. We show how to use the simulation and the iLQR algorithm to evaluate the robot design. Then we introduce our robot hardware and experiments.


\subsection{Robot Simulation}\label{sec_exp_sim}
We construct a Matlab Simulink model as shown in Figure \ref{fig:sec1_monkey} (b). The dynamic equation of the robot and its linearizations are also implemented for iLQR. We use Matlab symbolic toolbox to linearize the system dynamics. The trajectory optimization algorithm runs on a laptop with i7-8750H CPU and 32GB RAM. For all robot designs, when solving Problem \ref{eqn:problem}, we set $T = 2T_{freefall}$, where $T_{freefall}$ is the time that the right hand takes to reach the robot's lowest point with zero control. $T_{freefall}$ can be easily measured in Simulink. We set $\Delta T=T/300$, $Q = \text{diag}[0, 0, 0, 0.02, 0.02, 0.02]$, $R = \text{diag}[0.3, 0.3]$, and $Q_f=\text{diag}[6400, 6400, 6400, 1e-5, 1e-5, 1e-5]$. Most problems can be solved within 30 iLQR iterations in less than 60 seconds. 

\subsection{Design Parameter Selection}
The simulation and the iLQR algorithm give us a testbed for testing design parameters. Although we can choose the length and mass distribution of arms and the body of the simulated robot arbitrarily, how to choose values for components of the real robot are more challenging. In simulation we can evaluate different robot designs quickly since it is convenient to change model parameters, try different initial configurations, and add external force disturbances in Simulink. We set different parameter combinations in the simulation and examine the brachiation performances of the resultant motion by observing the final cost of iLQR algorithm. 

\begin{figure}
    \centering
    \includegraphics[width=\linewidth]{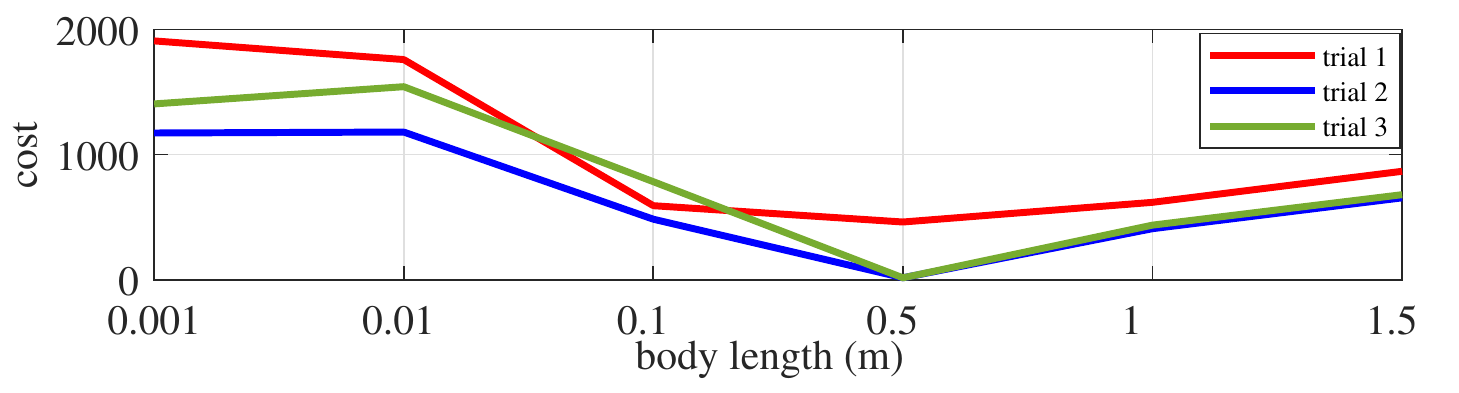}
    \caption{The effect of different body lengths
    on final cost. Three mass distributions are considered: Case 1) body mass = 3.5kg, arm mass = 2.025kg; Case 2) body mass = 3kg, arm mass = 3.025kg; Case 3) body mass = 3.0675kg, arm mass = 3.0675kg. In all cases body length between 0.1 - 0.5m gives the lowest cost. }
    \label{fig:sec_body_length}
\end{figure}

The first design decision that we consider is the necessity of having a body. To test this, we simulate different body designs consisting of a fixed total mass and arm length but variable mass distribution between the arms and body as well as variable body lengths. The results of these trials is shown in Fig. \ref{fig:sec_body_length}. For each mass distribution and body length we let the robot plan the brachiation trajectory and record the final cost of iLQR algorithm. It can be seen that for each mass distribution there exists an optimal body length that is greater than zero. The result aligns with the conclusion in biological studies that body momentum aids brachiation \cite{usherwood2003understanding}.

Another design decision we need to make is the ratio of the mass of the arm and the body. We refer to data presented in biology works as guidance. \cite{michilsens2011pendulum} reports parameters of siamangs based on anatomical measurement. It shows that on average one arm takes $9.1\%$  of total mass. \cite{bertram2001mechanical} presents data from a 7.95kg gibbon, whose arm has a mass of 0.95 (11.9\% of the total mass). We evaluate the arm and body ratio using iLQR and find that the lighter the arm is, the lower the cost will be. The result can be understood in an extreme case where the system becomes a pendulum with a massless rod attached to it. This system can reach every point in its end effector space without control effort. So the arm should be as light as possible. 

With these two major design decisions we can roughly form the general shape of the robot. Given the form factor of servo motors and available materials for the gripper hand, one arm is designed to have 0.38kg. The arm length is chosen to be 0.3m so the robot can travel along monkey bars that have distance within 0.3-0.4m. Then other robot parts can be designed accordingly. The following table summarizes key physical parameters of the robot we finally adopted: 
\begin{center}
 \begin{tabular}{||c | c||} 
 \hline
 \textbf{Item} &\textbf{Value}   \\ 
 \hline
 Arm length & 309.8 mm  \\  
 Arm mass & 0.384 kg \\ 
 Arm inertia (X) & 0.001694 kg$\cdot$m$^2$ \\   
 Arm inertia (Z) & 0.0002355 kg$\cdot$m$^2$\\  
 Body length & 81.82 mm\\ 
 Body mass & 2.111 kg\\ 
 Body inertia (X) & 0.01712 kg$\cdot$m$^2$ \\  
 Body inertia (Z) & 0.01230 kg$\cdot$m$^2$ \\ 
 \hline
 \end{tabular}
    
\end{center}

In the table, inertia values are measured in the CAD model. The proportion of the arm to the total mass is $13.3\%$, close to the animal data. For this robot, the swing time is chosen to be $T=0.66s$. The final cost solved by iLQR algorithm for swinging between monkey bars with distance 0.4m is 1.14, comparing to the initial cost 2640 using zero control input. The final control sequence gives a precise and low energy consumption trajectory.

\subsection{Robot Implementation}
\begin{figure}
    \centering
    \includegraphics[width=0.8\linewidth]{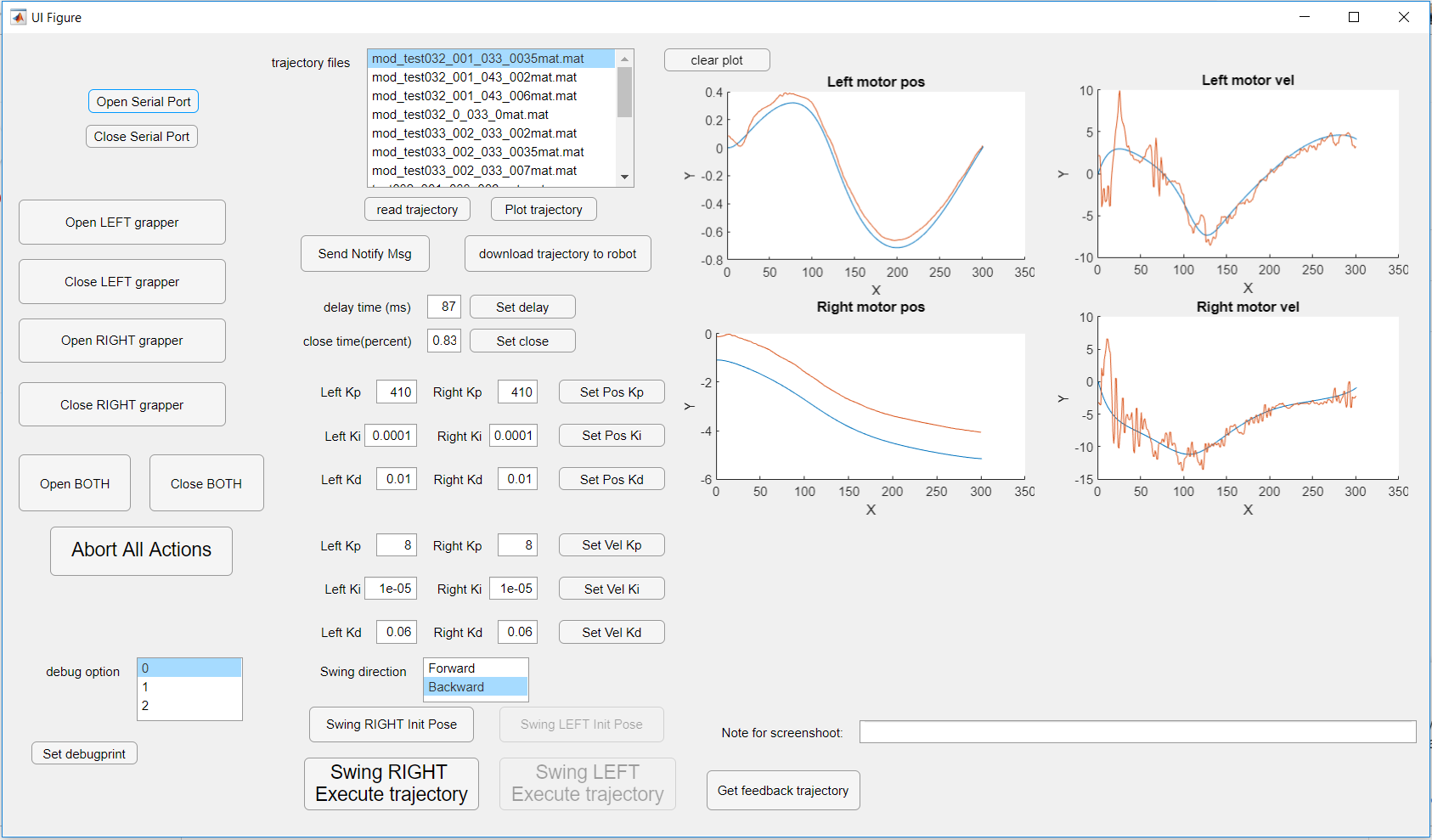}
    \caption{Matlab control GUI and plots of desired joint values and actual joint values.}
    \label{fig:sec_exp_plot}
\end{figure}

The brachiation robot we built is shown in Figure \ref{fig:sec1_monkey}. The final mass and length of components follow the design decisions made in the previous section.  
The shoulder motors are DJI RoboMaster GM6020 motor\footnote{DJI RoboMaster GM6020 Motor,  \url{https://www.robomaster.com/en-US/products/components/general/gm6020/info}}, which provide 1.2Nm continuous torque and 200rpm rotational velocity. Each arm has a gripper containing a four bar linkage driven by a PWM servo motor. The servo allows the gripper to close within 0.1 seconds. These motors are directly controlled by a STM32F427 MCU. The MCU estimates the robot state from motor encoders and an onboard IMU. Trajectory generation is computationally heavy for an MCU so we keep the trajectory generation algorithm on the laptop. The MCU and the laptop communicate via Bluetooth. A Matlab GUI is implemented to monitor and control communications. Figure \ref{fig:sec_exp_plot} shows the GUI and how does it plot the desired joint angles and velocities against actual joint angles and velocities. Users can click buttons to send computed trajectory to MCU, move the robot into the initial configuration $x_0$, and start the trajectory tracking control for one continuous brachiation cycle. 

\subsection{Experiments}
To demonstrate the brachiation ability of the robot, we built a test stand with monkey bars with adjustable distances. In experiments, we first hang the robot onto two bars, and then generate the trajectory according to the bar distances. Then we send the trajectory to the robot to execute the trajectory. The robot first moves to the initial configuration before each motion and then executes the swing. For multiple bars with the same distance, we reuse the same trajectory with additional logic to switch the role of left and right hand after each swing. For 0.3m even distance brachiation, the success rate is 100\% over 20 trials.
A video containing consecutive swings among evenly placed bars and swings between bars with different distances 
can be viewed at \href{https://youtu.be/Rvz2Wcv0qVQ}{https://youtu.be/Rvz2Wcv0qVQ}

%% file: sections/sec_conclude.tex

In this paper we present the design and implementation of a three-link brachiation robot. Modeling, planning, and control were all considered in the design. Experiments in simulation and on the real robot show that the trajectory generation and tracking controller are precise and effective. Moreover, the design process of using iLQR cost to evaluate robot design helped guide the design of the body and arms. This design methodology could potentially shed light on other dynamical robot designs as well.  

Our current optimal control based planning algorithm is limited to one single continuous swing, and so in future work we will study how to plan motion across multiple swings as well as consider ricochetal brachiation (which contains an air-borne phase). Hybrid dynamics and nonholonomic constraints will be required in these cases necessitating a new controller design.

    